# The Change of the Course of Time in a Force Field and Imponderability

**Oleinik V.P.**

Department of General and Theoretical Physics,
National Technical University of Ukraine **"**Kiev Polytechnic Institute**"**,
Prospect Pobedy 37, Kiev, 03056, Ukraine; e-mail: yuri@arepjev.relc.com
http://www.chronos.msu.ru/lab-kaf/Oleynik/eoleynik.html

**Abstract.** It is shown that the force in relativistic mechanics is not only the cause of acceleration of particle relative to an inertial frame of reference, but also the cause of change of the course of time along the particle's trajectory. Therein lies the physical content of the dynamical principle underlying the special theory of relativity (relativistic mechanics). The general formula for the relative course of time between the points lying on the trajectory of motion of particle under the action of a force field in an inertial reference frame is derived. The applications of the theory developed to homogeneous fields - to the field of gravity and electromagnetic field, and also to the gravitational field produced by a point mass particle are considered. Physical properties of the state of imponderability of particle in an external force field are investigated. It is noted that the change in the course of time in a force field is in no way connected with the change in space-time metric and is a direct consequence of the causality principle of relativistic mechanics.

## 1. Introduction

Time is among the most common concepts, which are used constantly both in everyday life and in science. This is because all the events and material processes in the world happen in space and develop in time and, hence, the laws that govern space-time connections are the most general and hold for all the forms of matter. Nevertheless, time remains one of the most mysterious concepts of physics; its physical essence is not adequately revealed up till now [1-4]. The concept of time with difficulty yields to logical analysis.

From the point of view of common sense the essence of time is that time characterizes the duration of events and processes, indicates their natural sequence, at which the present, going away to the past, gives place to the future.

I. Newton gave a clear-cut characteristic of the concept of time, to which the majority of physicists adheres: "The absolute, true, and mathematical time in itself and by virtue of its nature flows uniformly and regardless to any other object". Though, according to Newton, time flows equally and uniformly and does not depend on the processes, occurring in the world, the daily experience speaks in favour of the fact that the course of time is not uniform. Depending on circumstances in our history, it seems to us that time either flies swiftly or hangs heavy on our hands; sometimes it even changes suddenly, by leaps. In connection with these speculations the question arises of whether the subjective sensations of non-uniformity in the course of time familiar to everyone have an objective basis.

In Newtonian mechanics time is of an absolute character, it does not change as one passes from one inertial reference frame to another and represents merely a parameter, whose change at the will of explorer results in the change of state of a mechanical system in accordance with the equation of motion.

In relativistic mechanics time remains a parameter describing the development of system. But now time and space are intimately linked with each other to form a single whole – the 4-dimensional space-time. In going from one inertial frame of reference to another time gets entangled with spatial coordinates, so that time in one reference frame represents a "mixture" of time and coordinates in the other. Time ceases to be universal, the same in all inertial reference frames; it takes on a relative character.

The indissoluble association of time and space takes on special importance in the light of the concept of physical field, which was called by Einstein the most important discovery in physics after the times of Newton. According to this concept, the occurrence in space of a force field means that space turns into a physical environment, which is capable to interact directly with other bodies and gains, thus, physical properties, becoming an active participant of physical processes. In view of the fact that space and time are indissolubly related to each other, the presence of a force field in some area of space must necessarily result in the appearance of physical properties of time caused by the motion of body in this area.

Thus, from the synthesis of the notion of space-time and of the idea of physical field it follows with necessity that the course of time in a given region of space should depend on physical processes in this region, i.e. time, as well as space, should have physical properties [5-8].

It should be emphasized that in special theory of relativity (STR) time and spatial coordinates are independent and formally equal in rights quantities, which determine the position of elementary events in space-time. On the other hand, time stands out in relation to spatial coordinates. The special role of time is due, from the viewpoint of geometry, to the pseudoeuclidity of geometry of the 4-dimensional space. From the physical point of view, it is associated with the dynamical principle (causality principle), according to which the state of motion of a physical system at an instant of time $t$ uniquely defines its behaviour at the next instant of time $t + 0$. The significance of dynamical principle lies in the fact that it relates the temporal evolution of system to the physical processes caused by force fields and in doing so it allows one to determine the course of time in the system, its possible dependence upon the character of physical processes, and not just the sequence of events and their duration.

The idea about the existence of the physical properties of time belongs to N. Kozyrev [9]. By introducing into mechanics an additional parameter taking into account the directivity of the course of time, Kozyrev has formulated causal (asymmetrical) mechanics from which it follows that time has physical properties. According to the results of theoretical and experimental investigations conducted by Kozyrev and his followers [9-13], events can proceed not only in time, but also with the help of time, information being transmitted not through force fields, but via a temporal channel, and the transfer of information happens instantaneously. The appearance of additional forces, associated with the physical properties of time and capable to fulfill work, testifies that time can serve as a power source.

In the papers by I. Eganova [12] and M. Lavrent'ev and I. Eganova [13] the problem is stated of direct experimental research of the physical properties of time with the purpose to ascertain the relations of a new type between phenomena and to discover new methods of changing the state of substance. In [14] O. Jefimenko investigated the dynamical effect of the slowing-down of time.

According to [6-8], the conclusion that physical properties of time exist follows strictly from relativistic mechanics, without introducing any additional hypotheses. The physical properties of time are of purely dynamical nature: their existence results from dynamical principle. The availability of physical properties of time is manifested in that time has a local inhomogeneity: its course along the trajectory of motion of a point particle in a force field is continuously changed, and this change in the course of time is a result of the action upon the particle of a force field in the inertial reference frame, in which the motion is considered.

In view of fundamental importance of the problem considered, we shall discuss the physical content of the local dynamical inhomogeneity of time in more detail.

Let us consider the motion of a classical point particle under the action of a force field in the inertial reference frames $K$ and $K'$, moving relative to each other. The Cartesian coordinates connected with the reference frames are assumed for definiteness to be oriented in such a manner that the $x, y, z$-axes of the frame $K$ are parallel to the $x', y', z'$-axes of the frame $K'$, the $x$-axis and the $x'$-axis coincide with each other, and the reference frame $K'$ moves with a velocity $V_0$ relative to the $K$-frame along the $x$-axis. Denote by $dl_A$ the length of a path section in a vicinity of a point $A$ in the reference frame $K$, which the particle covers for the time $dt_A$, and by $dl'_A$ and $dt'_A$ the corresponding quantities relating to the reference system $K'$. The quantities $dt'_A$ and $dt_A$ are connected with each other by equality [6]:

$$dt'_A = \gamma \left( 1 - \frac{V_0\, u_x(t_A)}{c^2} \right) dt_A, \tag{1}$$

where $\gamma = \left( 1 - \dfrac{V_0^2}{c^2} \right)^{-1/2}$, $\quad u_x(t)$ is the $x$-component of the particle velocity $\boldsymbol{u}(t)$ in the $K$-frame at the instant of time $t$.

Quantity $\dfrac{dt'_A}{dt_A}$ characterizes the change in the course of time in the vicinity of point $A$ on the particle's trajectory in the reference frame $K'$ as compared with the reference frame $K$. As is seen from (1), if during some interval of time the $x$-component of velocity of a particle is changed ($u_x(t) \neq const$), on this time interval the relative course of time is changed as well ($\dfrac{dt'_A}{dt_A} \neq const$). If the particle moves on a path section uniformly and rectilinearly, i.e. by inertia ($\boldsymbol{u}(t) = const$), the relative course of time on the path section is remained constant: $\dfrac{dt'_A}{dt_A} = const$. Inasmuch as the change in the velocity of particle in an inertial reference frame is conditioned, according to the main postulate of classical mechanics, by the action on particle of a force connected with some physical field, hence, the force acting on particle is the reason of change in the course of time along the particle's trajectory.

According to (1), the gist of the phenomenon of local dynamical inhomogeneity of time is that the quantity $dt'_A$ depends not only on $dt_A$, but also on the particle's velocity in the vicinity of the point $A$. As the change in the particle's velocity is determined by the force influence on the particle, it follows from here that the **force** acting on a particle in some inertial reference frame **is the reason for change in the course of time along the particle's trajectory.** In other words, the dynamical inhomogeneity of time means that the different instants of time on the time axis prove to be physically non-equivalent when the particle moves in a force field.

Let's go over from point $A$ to some other point $B$, also lying on the particle's trajectory, and write down for it the relationship analogous to (1):

$$dt'_B = \gamma \left( 1 - \frac{V_0\, u_x(t_B)}{c^2} \right) dt_B.$$

By dividing last equality and equality (1) term by term, we arrive at relationship

$$\frac{dt_A}{dt_B} = \frac{\left( 1 - \dfrac{V_0\, u_x(t_B)}{c^2} \right) dt'_A}{\left( 1 - \dfrac{V_0\, u_x(t_A)}{c^2} \right) dt'_B}. \tag{2}$$

Quantities $\dfrac{dt_A}{dt_B}$ and $\dfrac{dt'_A}{dt'_B}$ characterize the relative course of time between the points $A$ and $B$ on the particle's trajectory in the reference frames $K$ и $K'$, respectively. In virtue of (2), if $u_x(t_A) = u_x(t_B)$, then $\dfrac{dt_A}{dt_B} = \dfrac{dt'_A}{dt'_B}$, i.e. the relative course of time between points $A$ and $B$ in the $K$-frame coincides with that in the $K'$-frame. To change the relative course of time between two points in one inertial reference frame as compared to the other, it is necessary that a force should act on the particle on the corresponding path section. According to (2), $dt_A \neq dt_B$ at $dt'_A = dt'_B$ if only the particle's velocities at points $A$ and $B$ are not identical in magnitude. In this case, if $dl'_A = dl'_B$, then, generally speaking, $dl_A \neq dl_B$, i.e. to equal distances, which the particle passes in different regions of space in reference frame $K'$, there correspond different distances traversed by the particle in reference frame $K$. This can be caused by both the different course of time at points $A$ and $B$ and the different velocities of particle at these points. In this connection the questions arise: How can these factors be separated from each other? How can the relative course of time be determined between the different points of space being considered in the same inertial reference frame?

Relying on the reasoning given above, it is natural to suppose that the change in the course of time at the points lying on the particle's trajectory can be caused only by the action of a force field on the particle. Indeed, in the absence of force field, when the particle moves by inertia, there are no reasons for changing the course of time. Let's go over from the inertial reference frame $K$, in which the motion of a particle takes place under the action of a force field, to such a noninertial reference frame $\tilde{K}'$, in which the inertial force is completely compensated for by the force field action at that point of space where the particle is at rest. Apparently, in the $\tilde{K}'$-frame the particle moves by inertia, i.e. it is in a free state (imponderability state) [15,16]. Since in this state the force effect on the particle is absent and, thus, the reason of changing the course of time is lacking, in this reference frame the course of time should be uniform at the point where the particle is placed. By choosing in the $\tilde{K}'$-frame at the point where the particle is placed two equal in magnitude intervals of time corresponding to two different points $A$ and $B$ lying on the particle's trajectory in the inertial reference frame $K$, and by performing then the inverse transition from the reference frame $\tilde{K}'$ to the $K$-frame, we may define the magnitude of relative course of time between points $A$ and $B$ in the reference frame $K$.

In this paper, on the basis of the reasoning above, the relative course of time is considered between the points lying on particle's trajectory in an inertial reference frame in uniform external field, and also in gravitational field created by a material point.

Note that equality (1) is one of the relationships entering into the Lorentz transformations for coordinates and time, and consequently it is not new. A new point is that an analysis of this relationship, as applied to the motion of point particle in a path under the action of a force field, is given and on its basis a number of physical consequences is derived concerning the course of time, which were not discussed in literature till now and were formulated for the first time in [5-8]. These consequences are important not only for elucidating the physical nature of time, but also for a deep insight into the physical content of relativistic mechanics, and thus they deserve consideration in more detail.

According to A. Logunov, the main content of special theory of relativity (STR) consists in that "all physical processes proceed in space-time possessing pseudoeuclidean geometry" ([4], p.26). Undeniably, from the mathematical (geometrical) point of view, in the formulation presented above the essence of STR is expressed correctly, still **the physical content of relativistic mechanics, the physical essence of dynamical principle underlying it, is that the force is not only the reason of particle's acceleration in an inertial reference frame, but also the reason of the change in the course of time along particle's trajectory.** It should be emphasized that existence of a link between the force and the course of time along particle's trajectory follows directly from the fact that space and time are connected with each other to form a single 4-dimensional space.

Thus, the fundamental difference between relativistic mechanics and Newtonian one lies not only in the fact that in Newtonian mechanics time is of an absolute character and does not change in going over from one inertial frame of reference to another, and in STR it ceases to be identical in all inertial reference frames. In relativistic mechanics time with necessity acquires physical properties, which are conditioned by the action on particle of a force connected with a physical field. As a result of the force action, the course of time is continuously changed along the particle's trajectory.

In connection with the phenomenon of dynamical inhomogeneity of time, let us consider equality

$$dx = \gamma \left( 1 + \frac{V_0}{u'_x(t')} \right) dx', \tag{3}$$

with $dx$ and $dx'$ being the increments of coordinates in the reference frames $K$ and $K'$. It would seem, equality (3) points to the existence of the dynamical inhomogeneity of spatial coordinates similar to the time inhomogeneity. Indeed, according to (3) the distance $dx$, traveled by a particle in the $K$-frame, depends not only on the distance $dx'$, which the particle passes in the $K'$-frame, but also on the instant of time $t'$. This dependence cannot be treated, however, as a manifestation of the dynamical inhomogeneity of spatial coordinates. Formula (3) is a trivial consequence of kinematics: it expresses the simple fact that the distance traveled by particle depends on its velocity, which may change with time (as a result of the action of force). Really, in classical mechanics, on the basis of Galileo transformations, we have successively:

$$dx = u_x \, dt = \left( u'_x + V_0 \right) dt, \quad dx' = u'_x \, dt.$$

From this it follows that

$$dx = \left( 1 + \frac{V_0}{u'_x} \right) dx'.$$

Up to the factor $\gamma$ last formula coincides with (3). In the case of relativistic mechanics we have analogously:

$$dx = u_x(t) \, dt, \quad dx' = u'_x(t') dt'.$$

From here, using the velocity addition law and Lorentz transformations for time, we arrive at formula (3).

Thus, as distinct from spatial coordinates, time in relativistic mechanics may be deformed (i.e. time scale may change) under the action of an external force. It should be emphasized that the active role of time in dynamics is due to the dynamical principle: the latter is formulated in terms of time, not of spatial coordinates.

Let us enumerate the main results presented in the subsequent sections of the paper.

In section 2 a general formula is given for the relative course of time between the points lying on particle's trajectory in a force field in an inertial reference frame. It is demonstrated that a consistency condition is fulfilled for the theory of the change of the course of time in a force field developed in the paper.

The application of general theory to homogeneous fields - to electric field and the field of gravity is considered in section 3. It is noted here that, because of relativistic corrections to the equation of motion of particle, the homogeneous field cannot be compensated for completely by inertial forces. The imponderability state of relativistic particle in homogeneous field possesses stability: if we remove the particle from the state of imponderability with the help of an external force and then leave it to its own devices, the particle returns to the imponderability state.

Section 4 is devoted to the investigation of the course of time and imponderability state in homogeneous electromagnetic field. The formula is deduced for the relative course of time along particle's trajectory in arbitrary homogeneous electromagnetic field in nonrelativistic approximation.

In section 5 the application of general theory is considered to the motion of particle in the gravitational field created by mass point. The quasiinertial reference frames freely falling on a force center in different radial directions are noted to be physically nonequivalent to each other.

In final section it is emphasized that research on the problem of time is of fundamental importance in science.

## 2. The course of time and the state of imponderability of particle

Let us transform the equation of motion of classical point particle of mass $m$ in an external force field $\boldsymbol{F} = \boldsymbol{F}(\boldsymbol{r}, \dot{\boldsymbol{r}}, t)$, written in an inertial reference frame $K$,

$$m\ddot{\boldsymbol{r}} = \boldsymbol{F}, \tag{4}$$

with $r = r(t)$ being the radius-vector of the particle at the instant of time $t$, to the reference frame $\tilde{K}'$ moving arbitrarily relative to the reference frame $K$. Denote by $r_0 = r_0(t)$ and $r' = r'(t)$ the radius-vector drawn from the origin of the $K$ - frame to the origin of the $\tilde{K}'$ - frame and the radius-vector of the particle in the $\tilde{K}'$ - frame, respectively. The $r$ and $r'$ are related by equality

$$r = r' + r_0(t). \tag{5}$$

Coordinate axes of the reference frame $\tilde{K}'$ are supposed for convenience to be parallel to the corresponding coordinate axes of the $K$ -frame. This will allow one to consider the coordinate unit vectors lying on axes of the frames $\tilde{K}'$ and $K$ to be not varied with time. From (5) follow the transformation laws for velocities and accelerations:

$$\dot{r} = \dot{r}' + \dot{r}_0(t) \quad \text{и} \quad \ddot{r} = \ddot{r}' + \ddot{r}_0(t). \tag{6}$$

Substituting (5) and (6) into (4), we arrive at the equation of motion of particle in the reference frame $\tilde{K}'$:

$$m\ddot{r}' = \widetilde{F}, \tag{7}$$

where

$$\widetilde{F} = F(r' + r_0(t),\ \dot{r}' + \dot{r}_0(t), t) - m\ddot{r}_0(t). \tag{8}$$

Note that inertial force $F_{in}$, $F_{in} = -m\ddot{r}_0(t)$, does not depend upon quantities $r'$ and $\dot{r}'$, which are the dynamical variables of the particle in the reference frame $\tilde{K}'$ and dependent only on time $t$. Consequently, the force $F_{in}$ represents a homogeneous field varying with a time.

The expression for the force $\widetilde{F}$, entering into the right-hand side of (7), can be expanded in a Taylor series in $r'$ and $\dot{r}'$. Assuming quantities $|r'|$ and $|\dot{r}'|$ to be small as compared to $|r_0(t)|$ and $|\dot{r}_0(t)|$, accordingly, and retaining only the terms of the first order in magnitude, we can obtain:

$$F(r' + r_0(t),\ \dot{r}' + \dot{r}_0(t), t) = F_0 + F_1, \tag{9}$$

where

$$F_0 = F(r_0(t),\ \dot{r}_0(t), t) \equiv F_0(t),$$

$$F_1 = (r'\vec{\nabla}_a + \dot{r}'\vec{\nabla}_b)F(a, b, t) \equiv F_1(r', \dot{r}', t) \quad \text{при} \quad a = r_0(t), b = \dot{r}_0(t).$$

Let us require that the following condition be fulfilled:

$$m\ddot{r}_0(t) = F_0(t). \tag{10}$$

With (9) and (10) the equation of motion (7) can be written as:

$$m\ddot{r}' = F_1. \tag{11}$$

It is seen from expression for $F_1$ that $F_1 = 0$ at $r' = 0$, $\dot{r}' = 0$. This means that the particle being at rest at the origin of the reference frame $\tilde{K}'$ is free: in the state $r' = 0$, $\dot{r}' = 0$ the force of inertia is completely compensated for by the external force $F$. If the force $F$ is the force of gravity, the state under study represents an imponderability state of particle [15,16]. This term (the imponderability state) we retain also in the case that $F$ describes an arbitrary force field.

It should be emphasized that though in the reference frame $\tilde{K}'$ the imponderability state of the particle is realized, the reference frame essentially differs from the inertial one. The fundamental distinction between them is that in inertial reference frame the particle, being free at one point of space, remains free at any other point (i.e. the force effect on particle is lacking in the whole space: $F \equiv 0$), whereas in the reference frame $\tilde{K}'$ the force effect on particle for arbitrary external force field $F(r, \dot{r}, t)$ is lacking only at point $r' = 0$, $\dot{r}' = 0$ (see (9)-(11)). However, one always can indicate such a region of space-time, in which the forces of inertia acting on particle in the $\tilde{K}'$-frame approximately compensate for the forces acting on it on the part of external force field, so that in this space-time region the $\tilde{K}'$-frame may be considered approximately as an inertial frame.

To circumstantiate, in what sense and with what accuracy the $\widetilde{K}'$-frame is the inertial one, we shall represent the external force, acting on the particle, as a sum of two components: the force $\boldsymbol{F}(\boldsymbol{r}, \dot{\boldsymbol{r}}, t) \equiv \boldsymbol{F}$, which further will be partly compensated for by inertia force, and an additional force $\boldsymbol{f}$, which for simplicity is taken as being constant ($\boldsymbol{f} = const$) and whose function consists in deflecting the particle from the state of imponderability. By transforming the equation of motion (4), in which the substitution $\boldsymbol{F} \to \boldsymbol{f} + \boldsymbol{F}$ is fulfilled, to the reference frame $\widetilde{K}'$, we obtain the following equation of motion (cf. (7) and (8)):

$$m\ddot{\boldsymbol{r}}' = \boldsymbol{f} + \widetilde{\boldsymbol{F}}. \tag{12}$$

In the region of space-time (we shall call it region $P$), in which the following condition is fulfilled

$$\left| \widetilde{\boldsymbol{F}} \right| << \left| \boldsymbol{f} \right|, \tag{13}$$

force $\widetilde{\boldsymbol{F}}$ in expression (12) can be neglected as compared with the force $\boldsymbol{f}$ and as a result we come to the equation $m\ddot{\boldsymbol{r}}' = \boldsymbol{f}$, according to which the acceleration of particle in the $\widetilde{K}'$-frame is conditioned only by the action of external force $\boldsymbol{f}$, as it should be for a true inertial reference frame. Consequently, in describing the motion of particle in the region $P$ the reference frame $\widetilde{K}'$ can be considered as inertial one. Apparently, the region $P$ lies in a vicinity of the point, at which

$$\widetilde{\boldsymbol{F}} = 0, \tag{14}$$

and the accuracy, with which the frame $\widetilde{K}'$ is inertial, is determined by the accuracy, with which inequality (13) is fulfilled.

Thus, in describing the motion in the noninertial reference frame, in which the imponderability state of particle is achieved, i.e. equality (14) holds, there exists such a space-time region in which this reference frame differs very little from the inertial one. For this reason it is natural to call such a reference frame the quasiinertial one. Obviously, there is infinitely many quasiinertial reference frames physically equivalent to each other. In particular, the quasiinertial reference frames moving uniformly and rectilinearly relative to each other are physically equivalent, if only the motion of particle is considered in the space-time regions, in which the force $\widetilde{\boldsymbol{F}}$ can be neglected as compared with the external force $\boldsymbol{f}$. On the other hand, if we want to take into consideration corrections to the solution of equation (12) conditioned by the force $\widetilde{\boldsymbol{F}}$, then the quasiinertial reference frames mentioned above can be found to be physically nonequivalent to each other. This is due to the fact that because of the vectors $\boldsymbol{r}_0$ and $\dot{\boldsymbol{r}}_0$ entering into equation (12) there can appear in space the preferential regions and directions.

The results presented above can easily be generalized to the relativistic case. We proceed from the equation of motion

$$\frac{d\boldsymbol{p}}{dt} = \boldsymbol{F}, \tag{15}$$

where

$$\boldsymbol{p} = m\boldsymbol{u}, \quad m = m_0\left(1 - \frac{u^2}{c^2}\right)^{-\frac{1}{2}},$$

$\boldsymbol{u} = \dot{\boldsymbol{r}}$, $\boldsymbol{r} = \boldsymbol{r}(t)$, $m_0$ is the rest mass of particle. With the aid of equality

$$\frac{d}{dt}mc^2 = \boldsymbol{u}\boldsymbol{F}, \tag{16}$$

which follows from (15), relativistic equation of motion can be recast in the quasiclassical form [4] convenient for further analysis:

$$m_0\frac{d\boldsymbol{u}}{dt} = \left[\boldsymbol{F} - \frac{\boldsymbol{u}(\boldsymbol{u}\boldsymbol{F})}{c^2}\right]\left(1 - \frac{u^2}{c^2}\right)^{\frac{1}{2}}. \tag{17}$$

As is clarified in [4], in order to describe the motion of particle within the framework of relativistic mechanics we have a right to use both inertial and noninertial reference frames. At transition from an inertial reference frame to the noninertial one the space-time geometry is not changed and remains pseudoeuclidean. Passing on to the noninertial reference frame $\tilde{K}'$, we shall take advantage of equalities (5) and (6) and introduce notation:

$$\boldsymbol{u} = \dot{\boldsymbol{r}}, \quad \boldsymbol{u}' = \dot{\boldsymbol{r}}', \quad \boldsymbol{u}_0 = \dot{\boldsymbol{r}}_0(t).$$

Further we substitute the first of equalities (6) in (17) and, assuming the conditions

$$|\boldsymbol{u}'| << |\boldsymbol{u}_0|, \quad \frac{|\boldsymbol{u}'\boldsymbol{u}_0|}{c^2} << 1 - \frac{u_0^2}{c^2} \tag{18}$$

to be fulfilled, expand the right-hand side of equation (17) in a series in powers of $\boldsymbol{r}'$ and $\boldsymbol{u}'$. Retaining in the expansion only terms of the first order, we arrive at the following equation of motion for particle in the $\tilde{K}'$-frame:

$$m_0 \frac{d\boldsymbol{u}'}{dt} = -\left( \frac{d\boldsymbol{p}_0}{dt} - \boldsymbol{F}_0 \right)\left( 1 - \frac{u_0^2}{c^2} \right)^{1/2} +$$

$$+ \left[ \boldsymbol{F}_1 - \frac{\boldsymbol{u}_0(\boldsymbol{u}_0\boldsymbol{F}_1)}{c^2} - \frac{\boldsymbol{u}'(\boldsymbol{u}_0\boldsymbol{F}_0)}{c^2} - \frac{\boldsymbol{u}_0(\boldsymbol{u}'\boldsymbol{F}_0)}{c^2} \right]\left( 1 - \frac{u_0^2}{c^2} \right)^{1/2} - \tag{19}$$

$$- \left[ \boldsymbol{F}_0 - \frac{\boldsymbol{u}_0(\boldsymbol{u}_0\boldsymbol{F}_0)}{c^2} \right]\frac{\boldsymbol{u}'\boldsymbol{u}_0}{c^2}\left( 1 - \frac{u_0^2}{c^2} \right)^{-1/2}.$$

Here expansion (9) and notation $\boldsymbol{p}_0 = m_0\boldsymbol{u}_0\left( 1 - \dfrac{u_0^2}{c^2} \right)^{-1/2}$ are used. Quantity $\boldsymbol{r}_0(t)$ is determined by equation

$$\frac{d\boldsymbol{p}_0}{dt} = \boldsymbol{F}_0. \tag{20}$$

Since in this case the right-hand side of equation (19) vanishes at $\boldsymbol{r}' = 0$, $\boldsymbol{u}' = 0$, the particle, which is at rest at the origin of the reference frame $\tilde{K}'$, turns out to be free. At $\boldsymbol{r}' \neq 0$, $\boldsymbol{u}' \neq 0$ a force arises acting on particle. Thus, the state of particle with $\boldsymbol{r}' = 0$, $\boldsymbol{u}' = 0$ turns out to be the imponderability state.

In the inertial reference frame $K$, described by Galilean coordinates $t, x, y, z$, the square of space-time interval is of the form $ds^2 = c^2dt^2 - dx^2 - dy^2 - dz^2$. From here it follows that coordinate $t$ has the meaning of physical time, and the rest of coordinates determine space intervals along corresponding axes. At the same time, in the reference frame $\tilde{K}'$ coordinate $t$ represents a coordinate time and has no physical meaning. To obtain physical time $d\tau$ and physical length $dl$ in the $\tilde{K}'$-frame, let us transform the square of space-time interval $ds^2 = c^2dt^2 - d\boldsymbol{r}^2$ to the frame $\tilde{K}'$, using the first of the equalities (6), and then separate out in the expression obtained the full square containing time coordinate. We have successively:

$$ds^2 = c^2dt^2 - \left( d\boldsymbol{r}' + \boldsymbol{u}_0 dt \right)^2 = \left( \sqrt{c^2 - u_0^2}\, dt - \frac{\boldsymbol{u}_0 d\boldsymbol{r}'}{\sqrt{c^2 - u_0^2}} \right)^2 - \left( d\boldsymbol{r}' \right)^2 - \left( \frac{\boldsymbol{u}_0 d\boldsymbol{r}'}{\sqrt{c^2 - u_0^2}} \right)^2.$$

Putting

$$ds^2 = c^2(d\tau)^2 - (dl)^2$$

and comparing the right-hand sides of two last relationships with each other, one can derive:

$$d\tau = \frac{1}{c}\sqrt{c^2 - u_0^2}\, dt - \frac{\boldsymbol{u}_0 d\boldsymbol{r}'}{c\sqrt{c^2 - u_0^2}}, \quad (dl)^2 = (d\boldsymbol{r}')^2 + \left(\frac{\boldsymbol{u}_0 d\boldsymbol{r}'}{\sqrt{c^2 - u_0^2}}\right)^2. \qquad (21)$$

If the particle, being at rest at the origin of the frame of reference, is in the imponderability state, it is reasonable to consider its proper time to flow uniformly, i.e. the course of time of the particle at different instants of proper time to be identical. Really, as it was noted in the previous section, in the imponderability state there is no force action on the particle and, thus, there is no reason for changing the course of time at the point where the particle is located. Denote by $A$ and $B$ any two points lying on the particle's trajectory in the inertial reference frame $K$, which the particle passes at the instances of time $t_A$ and $t_B$. Let $d\tau_A$ and $d\tau_B$ be the intervals of proper time of the particle, being at rest at the origin of the $\tilde{K}'$-frame, corresponding to the intervals $dt_A$ and $dt_B$, during which the particle moves in a path in the reference frame $K$ in a vicinity of points $A$ and $B$ (it is assumed the instants of time $t_A$ and $t_B$ to lie on intervals $dt_A$ and $dt_B$, and the instants of proper time $\tau_A$ and $\tau_B$ to correspond to these instants). Using the first of the relationships (21) and taking into account that for the particle at rest $d\boldsymbol{r}' = 0$, the equality of the intervals of proper time $d\tau_A = d\tau_B$ can be rewritten as

$$\sqrt{1 - \frac{\boldsymbol{u}_0^2(t_A)}{c^2}}\, dt_A = \sqrt{1 - \frac{\boldsymbol{u}_0^2(t_B)}{c^2}}\, dt_B.$$

From the equality above the relative course of time can be found between points $A$ and $B$ on the particle's trajectory in the inertial reference frame $K$:

$$\frac{dt_A}{dt_B} = \sqrt{\frac{c^2 - \boldsymbol{u}_0^2(t_B)}{c^2 - \boldsymbol{u}_0^2(t_A)}}. \qquad (22)$$

Note that since we consider here the state of particle with $\boldsymbol{r}' = 0$, $\boldsymbol{u}' = 0$, in virtue of (6) $\boldsymbol{u}_0(t) = \boldsymbol{u}$, i.e. $\boldsymbol{u}_0(t)$ in (22) is the velocity of particle at the instant of time $t$ relative to the inertial reference frame $K$.

It should be emphasized that quantities $dt_A$ and $dt_B$ in (22) have no sense of time intervals during which the particle passes identical distances in the vicinity of points $A$ and $B$. The quantities above have the following meaning: they are those time intervals, which correspond to the identical intervals of proper time of the particle being in the imponderability state.

The relative course of time between the points above in an inertial reference frame $K'$ can be written in the form similar to (22):

$$\frac{dt_A'}{dt_B'} = \sqrt{\frac{c^2 - \boldsymbol{u}_0'^2(t_B')}{c^2 - \boldsymbol{u}_0'^2(t_A')}}, \qquad (23)$$

$\boldsymbol{u}_0'(t')$ being the velocity of particle at instant $t'$ in the reference frame $K'$. Substituting (22) and (23) in (2) and taking into account that $u_x(t) = u_{0x}(t)$, we obtain relationship

$$\sqrt{\frac{c^2 - \boldsymbol{u}_0^2(t_B)}{c^2 - \boldsymbol{u}_0^2(t_A)}} = \frac{\left(1 - \frac{V_0 u_{0x}(t_B)}{c^2}\right)}{\left(1 - \frac{V_0 u_{0x}(t_A)}{c^2}\right)}\sqrt{\frac{c^2 - \boldsymbol{u}_0'^2(t_B')}{c^2 - \boldsymbol{u}_0'^2(t_A')}}, \qquad (24)$$

which represents a consistency condition for the theory developed here. Condition (24) can be verified as being an identity whose validity follows from the known equality (see [4], p.61)

$$1 - \frac{u_0'^2}{c^2} = \frac{1 - \dfrac{V_0^2}{c^2}}{\left(1 - \dfrac{V_0 u_{0x}}{c^2}\right)^2}\left(1 - \frac{u_0^2}{c^2}\right),$$

with $u_0$ and $u_0'$ being the velocities of particle in the reference frames $K$ и $K'$, respectively. The fulfillment of consistency condition (24) is an important argument in favour of the theory developed which establishes a link between the force effect on particle and the course of time along its trajectory.

Expression (22) can be represented in the following equivalent forms:

$$\frac{dt_A}{dt_B} = \frac{1 + \dfrac{\varepsilon_{kin}(t_A)}{m_0 c^2}}{1 + \dfrac{\varepsilon_{kin}(t_B)}{m_0 c^2}} = \frac{1 + \dfrac{U(t_0) - U(t_A)}{m_0 c^2}}{1 + \dfrac{U(t_0) - U(t_B)}{m_0 c^2}}, \qquad (25)$$

where $\varepsilon_{kin}(t)$ and $U(t)$ are the kinetic and potential energies of particle at the instant of time $t$, provided that equality $\varepsilon_{kin}(t_0) = 0$ is fulfilled. As is seen from (22), if the particle's velocity is small as contrasted to the speed of light, the change of the course of time is a relativistically small quantity of the order of $\left(\dfrac{u_0}{c}\right)^2$. Retaining in the expansion of the right hand side of (25) only the main in magnitude term, one can obtain the following nonrelativistic formula:

$$\frac{dt_A}{dt_B} = 1 + \frac{1}{m_0 c^2}\left(\varepsilon_{kin}(t_A) - \varepsilon_{kin}(t_B)\right) = 1 - \frac{1}{m_0 c^2}\left(U(t_A) - U(t_B)\right). \qquad (26)$$

### 3. Uniform electric (gravitation) field

As an application of the results received, we first consider the motion of a particle in an external uniform field $F = const$. In this case $F_0 = F$, $F_1 = 0$ (see (9)) and therefore, according to (10) and (11), in the reference frame $K$ the particle moves with constant acceleration

$$a_0 = \frac{F}{m}, \qquad (27)$$

while in the reference frame $\tilde{K}'$ it is free, because in the last frame of reference the inertial force is completely compensated for by the external force $F$. From the solution of equation (11) at $F_1 = 0$, which is of the form $r' = a + bt$, with $a$ and $b$ being arbitrary constant vectors, it follows that the particle moving uniformly and rectilinearly in the reference frame $\tilde{K}'$ is in the imponderability state. Thus, the motion of the particle of mass $m$ relative to an inertial reference frame under the action of a uniform field $F$ is equivalent to the motion of a free particle relative to the noninertial reference frame $\tilde{K}'$, which moves relative to the inertial frame with acceleration (27). Transition to the noninertial frame of reference allows one to completely exclude from consideration the uniform force field at once in the whole space. It should be remarked that uniform field has no physical sense, since such a field does not exist in nature. The real physical fields describing interaction between particles are essentially nonuniform.

Passing on to the relativistic case, we shall consider the solution of equation (20) obeying the initial condition $p_0 = 0$ at $t = 0$: $p_0 = Ft$. The sought-for solution is given by

$$\boldsymbol{u}_0 = \boldsymbol{a}_0 t \left(1 + \left(\frac{a_0 t}{c}\right)^2\right)^{-\frac{1}{2}}, \quad \boldsymbol{a}_0 = \frac{\boldsymbol{F}}{m_0}. \tag{28}$$

The integration of last equation yields:

$$\boldsymbol{r}_0(t) = \boldsymbol{r}_0 + \boldsymbol{a}_0 \frac{c^2}{a_0^2}\left(\sqrt{1 + \left(\frac{a_0 t}{c}\right)^2} - 1\right), \quad \boldsymbol{r}_0 = \boldsymbol{r}_0(0). \tag{29}$$

Taking into account equation (20) and equality $\boldsymbol{F}_1 = 0$, the equation of motion (19) can be represented in the form:

$$m_0 \frac{d\boldsymbol{u}'}{dt} = -\frac{1}{c^2}\left[\boldsymbol{u}'(\boldsymbol{u}_0 \boldsymbol{F}) + \boldsymbol{u}_0(\boldsymbol{u}'\boldsymbol{F})\right]\left(1 - \frac{u_0^2}{c^2}\right)^{\frac{1}{2}} -$$
$$- \left[\boldsymbol{F} - \frac{\boldsymbol{u}_0(\boldsymbol{u}_0 \boldsymbol{F})}{c^2}\right]\frac{\boldsymbol{u}'\boldsymbol{u}_0}{c^2}\left(1 - \frac{u_0^2}{c^2}\right)^{-\frac{1}{2}}. \tag{30}$$

Comparing equation (30) with the corresponding nonrelativistic equation (11), one can see that in the relativistic case the transition to the noninertial reference frame does not remove completely uniform force field in the whole space. The force field enters into the right-hand side of equation (30) both in the explicit form and implicitly, through the vector $\boldsymbol{u}_0$ (see (28)). According to (30), the uniform external field cannot be compensated for completely by inertial forces because of relativistic corrections to the equation of motion. Note that for uniform field the force acting on particle in the reference frame $\tilde{K}'$ does not depend on the radius-vector of particle $\boldsymbol{r}'$, but depends on its velocity $\boldsymbol{u}'$. This means that the particle, being at rest in the reference frame $\tilde{K}'$ at arbitrary point of space, is in the imponderability state. The force effect on the particle arises, however, if we impart an initial velocity $\boldsymbol{u}' \neq 0$ to it, thereby deflecting the particle from the imponderability state.

To establish the character of motion of the particle deflected from the imponderability state, let us consider the following initial state:

$$\boldsymbol{r}'(0) = \boldsymbol{r}_0' \neq 0, \quad \boldsymbol{u}'(0) = \boldsymbol{u}_0' \neq 0. \tag{31}$$

Using expressions (28) and (29), we can derive from (5) and (6) the following initial conditions:

$$\boldsymbol{r}(0) = \boldsymbol{r}_0' + \boldsymbol{r}_0, \quad \boldsymbol{u}(0) = \boldsymbol{u}_0'. \tag{32}$$

The solution to equation of motion (15) subject to initial conditions (32) can be written in the form:

$$\boldsymbol{p}(t) = \boldsymbol{F}t + m_0 \boldsymbol{u}_0'\left(1 - \frac{u_0'^2}{c^2}\right)^{-\frac{1}{2}}, \quad \boldsymbol{u}(t) = \frac{\boldsymbol{p}(t)}{m_0}\left(1 + \left(\frac{\boldsymbol{p}(t)}{m_0 c}\right)^2\right)^{-\frac{1}{2}}. \tag{33}$$

Taking into account (5), (6), (28) and (33), we can find:

$$\boldsymbol{u}'(t) = \boldsymbol{u}(t) - \boldsymbol{u}_0(t) \to 0 \quad \text{at} \quad t \to \infty. \tag{34}$$

Next, we use the energy conservation law resulting from equation (15) (see (16)):

$$m_0 c^2\left(1 - \frac{\boldsymbol{u}^2(t)}{c^2}\right)^{-\frac{1}{2}} - \boldsymbol{F}\boldsymbol{r}(t) = const. \tag{35}$$

Here the constant will be determined from the initial conditions (32):

$$m_0 c^2\left(1 - \frac{\boldsymbol{u}_0'^2}{c^2}\right)^{-\frac{1}{2}} - \boldsymbol{F}\left(\boldsymbol{r}_0' + \boldsymbol{r}_0\right) = C_1 = const. \tag{36}$$

On the other hand, using (34), from (35) and (36) we can derive:

$$m_0 c^2\left(1 - \frac{\boldsymbol{u}_0^2(t)}{c^2}\right)^{-\frac{1}{2}} - \boldsymbol{F}\left(\boldsymbol{r}'(t) + \boldsymbol{r}_0(t)\right) = C_1 \quad \text{at} \quad t \to \infty.$$

Forasmuch as $\boldsymbol{r}_0(t)$ obeys the equation (20), the following conservation law holds true (cf. (35))

$$m_0 c^2 \left(1 - \frac{\boldsymbol{u}_0^2(t)}{c^2}\right)^{-\frac{1}{2}} - \boldsymbol{F}\boldsymbol{r}_0(t) = C_2 = const. \qquad (37)$$

It is seen from two last equations that

$$\boldsymbol{F}\boldsymbol{r}'(t) = C_2 - C_1 \neq 0 \quad \text{at} \quad t \to \infty. \qquad (38)$$

Combining (37) at $t = 0$ and (36), we can obtain:

$$C_2 - C_1 = m_0 c^2 \left(1 - \left(1 - \frac{\boldsymbol{u}_0'^2}{c^2}\right)^{-\frac{1}{2}}\right) + \boldsymbol{F}\boldsymbol{r}_0'.$$

From here and from (38) it follows that

$$\boldsymbol{F}\left(\boldsymbol{r}'(t) - \boldsymbol{r}_0'\right) = m_0 c^2 \left(1 - \left(1 - \frac{\boldsymbol{u}_0'^2}{c^2}\right)^{-\frac{1}{2}}\right) \neq 0 \quad \text{at} \quad t \to \infty.$$

Consequently, $\boldsymbol{r}'(t) = const$, $\boldsymbol{r}'(t) \neq \boldsymbol{r}_0'$ at $t \to \infty$. Thus, in homogeneous force field the state of particle $\boldsymbol{r}'(0) \neq 0$, $\boldsymbol{u}'(0) \neq 0$ evolves into an imponderability state:

$$\boldsymbol{r}'(t) = const, \quad \boldsymbol{u}'(t) = 0 \quad \text{at} \quad t \to \infty \quad .$$

In other words, the imponderability state is characterized by stability: if we deflect a particle from an imponderability state with the help of some external force and then leave it on its own, the particle returns to a free state (though, generally speaking, $\boldsymbol{r}'(\infty) \neq \boldsymbol{r}'(0)$). Note that the conclusion made above concerning the stability of imponderability state of particle in homogeneous field is precise: in obtaining it no approximations based on the expansion of the right-hand side of equation (17) in powers of $\boldsymbol{r}'$ and $\boldsymbol{u}'$ were used.

The relative course of time between points $A$ and $B$, lying on the trajectory of motion of particle in homogeneous field in an inertial reference frame, is calculated from formulas (22) and (28) to yield:

$$\frac{dt_A}{dt_B} = \sqrt{\frac{1 + \left(\frac{a_0 t_A}{c}\right)^2}{1 + \left(\frac{a_0 t_B}{c}\right)^2}} . \qquad (39)$$

In the nonrelativistic approximation we obtain:

$$\frac{dt_A}{dt_B} = 1 + \frac{a_0^2}{2c^2}(t_A^2 - t_B^2). \qquad (40)$$

According to (39), if $t_B > t_A$, then $dt_B > dt_A$, i.e. when a particle moves in homogeneous field, the relative course of time along the particle's trajectory is increased: with increasing time coordinate in inertial reference frame time flows faster at the point where the particle is situated.

As a homogeneous field, let us consider the field of gravity force $\boldsymbol{F} = m_0 \boldsymbol{g}$, where $\boldsymbol{g} = const$ is the free fall acceleration of particle. By putting $\boldsymbol{g} = -g\boldsymbol{e}_z$, where $\boldsymbol{e}_z$ is the unit vector directed along the $z$-axis of the inertial reference frame $K$, we can express the force $\boldsymbol{F}$ through potential $\varphi$:

$$\boldsymbol{F} = -m_0 \vec{\nabla}\varphi, \quad \varphi = gz + const. \qquad (41)$$

Formulas (28) and (29) with $\boldsymbol{a}_0 = \boldsymbol{g}$ describe the free fall of particle with vanishing initial velocity.

In virtue of (29), at $\frac{gt}{c} \ll 1$ the coordinate of a freely falling particle in the $K$-frame can be written in the form:

$$z = z_0 - \frac{1}{2} g t^2, \quad z_0 = \text{const}. \tag{42}$$

As is seen from (39)-(42), at the free fall of particle relative to inertial reference frame, the field potential decreases and time flows faster and faster at the point of particle's localization. Calculate the proper time interval with the aid of formula (21) (at $dr' = 0$) by assuming the condition $\frac{gt}{c} \ll 1$ to be fulfilled. Taking into account (28), (41) и (42), we obtain:

$$d\tau = \sqrt{1 - \frac{\boldsymbol{u}^2(t)}{c^2}} dt = \left(1 - \frac{1}{2}\left(\frac{gt}{c}\right)^2\right) dt = \left(1 + \frac{\varphi}{c^2}\right) dt. \tag{43}$$

In deriving the last formula the constant in (41) was defined from the condition for the vanishing of potential $\varphi$ at $t = 0$. The relative course of time in inertial reference frame between points $A$ and $B$, which can be determined by equality $d\tau_A = d\tau_B$, is expressed by (40) with $a_0 = g$.

## 4. Homogeneous magnetic and electric fields

At first we dwell upon a homogeneous magnetic field $\boldsymbol{B} = \text{const}$. When a particle with charge $e$ moves in this field, the components of force $\boldsymbol{F}_0$ and $\boldsymbol{F}_1$ (see (9)) are given by:

$$\boldsymbol{F}_0 = \frac{e}{c}[\boldsymbol{u}_0 \boldsymbol{B}], \quad \boldsymbol{F}_1 = \frac{e}{c}[\boldsymbol{u}' \boldsymbol{B}].$$

Equations (10) and (11) can be written in the form

$$\ddot{\boldsymbol{r}}_0 = [\boldsymbol{\omega}\, \boldsymbol{u}_0], \tag{44}$$

$$\ddot{\boldsymbol{r}}' = [\boldsymbol{\omega}\, \boldsymbol{u}'], \tag{45}$$

where $\boldsymbol{\omega} = -\dfrac{e\boldsymbol{B}}{mc}$. According to (44) and (45), in the nonrelativistic case the transition to the noninertial reference frame $\widetilde{K}'$, whose origin rotates at angular velocity $\boldsymbol{\omega}$ about on axis passing through the origin of the $K$-frame, does not exclude the homogeneous magnetic field from equation of motion. Furthermore, the equation of motion of particle in the reference frame $\widetilde{K}'$ (45) does not differ in form from equation (44). As is seen from (45), the state of particle with

$$\boldsymbol{r}' = \boldsymbol{r}_0' + \boldsymbol{u}_{\mathrm{II}}' t \quad \text{at} \quad \boldsymbol{r}_0' = \text{const}, \quad \boldsymbol{u}_{\mathrm{II}}' = \text{const}, \quad \boldsymbol{u}_{\mathrm{II}}' \mid\mid \boldsymbol{\omega} \tag{46}$$

is an imponderability state. For the conservation law $\boldsymbol{u}'^2 = \text{const}$ is fulfilled, if we deflect the particle from the imponderability state, having imparted an initial velocity $\boldsymbol{u}_{0\perp}' \neq 0$ ( $\boldsymbol{u}_{0\perp}' \perp \boldsymbol{\omega}$ ) to it, and leave to its own devices, the particle will not return to the imponderability state.

The equation of motion of particle in the reference frame $\widetilde{K}'$ (19) differs from equation (45) by relativistic corrections. Otherwise the characteristics of motion in the reference frame $\widetilde{K}'$ in relativistic and nonrelativistic cases do not differ from each other.

For homogeneous magnetic field the conservation law $\boldsymbol{u}_0^2 = \text{const}$ follows from equation (20). Therefore in virtue of (22) $dt_A = dt_B$, i.e. the course of time along the particle's trajectory in homogeneous magnetic field in inertial reference frame is uniform.

We now turn to the discussion of homogeneous electric $\boldsymbol{E}$ and magnetic $\boldsymbol{B}$ fields, supposing for definiteness the field $\boldsymbol{B}$ to be directed along the $z$-axis. For a charged particle with charge $e$ the components of force $\boldsymbol{F}_0$ and $\boldsymbol{F}_1$ (see (9)) are given by

$$\boldsymbol{F}_0 = e\boldsymbol{E} + \frac{e}{c}[\boldsymbol{u}_0 \boldsymbol{B}], \quad \boldsymbol{F}_1 = \frac{e}{c}[\boldsymbol{u}' \boldsymbol{B}] \tag{47}$$

and therefore the equation of motion of particle in the reference frame $\widetilde{K}'$ can be written in the form (45). From here it follows that, as with homogeneous magnetic field, the state (46) is an imponderability state. The solution to equation (10), in which $\boldsymbol{F}_0$ is given by the first of the formulas (47), can be written as:

$$\boldsymbol{u}_0(t) = \left[\boldsymbol{\omega}\,\boldsymbol{r}_{0\perp}(t)\right] + \frac{e\boldsymbol{E}_{\mathrm{II}}}{m}t + \boldsymbol{u}_{0\mathrm{II}} + \frac{e}{m\omega^2}\left[\boldsymbol{\omega}\,\boldsymbol{E}_\perp\right]\,, \qquad (48)$$

where

$$\boldsymbol{r}_{0\perp}(t) = a\big(\sin(\omega\,t + \varphi_0),\,\cos(\omega\,t + \varphi_0),\,0\big), \quad \boldsymbol{E} = \boldsymbol{E}_{\mathrm{II}} + \boldsymbol{E}_\perp,$$

$$\boldsymbol{E}_{\mathrm{II}} = (0,0,E_z), \quad \boldsymbol{E}_\perp \perp \boldsymbol{\omega}, \quad \boldsymbol{u}_{0\mathrm{II}} = (0,0,u_{0z}), \quad \omega = \frac{eB}{mc},$$

$a$, $\varphi_0$ and $u_{0z}$ are arbitrary constants. From here we obtain the following formula for the kinetic energy of particle:

$$\varepsilon_{kin}(t) = \frac{m\,\boldsymbol{u}_0^2(t)}{2} = \frac{m}{2}\left\{\left(\boldsymbol{\omega}\,\boldsymbol{r}_{0\perp}(t) + \frac{e}{m\omega}\boldsymbol{E}_\perp\right)^2 + \left(\frac{eE_z}{m}t + u_{0z}\right)^2\right\}. \qquad (49)$$

Formula (26), in which $\varepsilon_{kin}(t)$ is given by expression (49), determines the relative course of time in an arbitrary constant in time homogeneous external field. In particular, in the case of crossed field ($E_z = 0$) the following expression is obtained ($m_0 = m$):

$$\frac{dt_A}{dt_B} = 1 + \frac{e}{mc^2}\,\boldsymbol{E}_\perp\big(\boldsymbol{r}_{0\perp}(t_A) - \boldsymbol{r}_{0\perp}(t_B)\big)\,.$$

For relativistic particle we restrict our consideration to the case of the fields $\boldsymbol{E}$ and $\boldsymbol{B}$ parallel to the $z$-axis. The solution of equation (20), in which function $\boldsymbol{F}_0$ is determined by the first of the formulas (47), can be written in the form [17]:

$$\begin{aligned} p_{0x} &= p_{0\perp}\,cos\,\varphi(t\,), \\ p_{0y} &= -p_{0\perp}\,sin\,\varphi(t\,), \\ p_{0z} &= eEt + \tilde{p}_{0z}, \end{aligned} \qquad (50)$$

where $p_{0\perp} = const$, $\tilde{p}_{oz} = const$, the function $\varphi = \varphi(t\,)$ obeys the equation

$$\frac{d\varphi}{dt} = \omega_0(t\,), \quad \omega_0(t\,) = \frac{eB}{mc}, \quad m = \sqrt{m_0^2 + \left(\frac{\boldsymbol{p}}{c}\right)^2}\,. \qquad (51)$$

The relative course of time can be calculated by formula (22), taking into account equalities (50) and (51) and relationship

$$1 - \frac{\boldsymbol{u}^2}{c^2} = \left(1 + \frac{\boldsymbol{p}^2}{(m_0 c)^2}\right)^{-1}\,.$$

The final result is:

$$\frac{dt_A}{dt_B} = \sqrt{\frac{1 + \dfrac{p_{0\perp}^2 + (eEt_A + \tilde{p}_{0z})^2}{(m_0 c)^2}}{1 + \dfrac{p_{0\perp}^2 + (eEt_B + \tilde{p}_{0z})^2}{(m_0 c)^2}}}\,. \qquad (52)$$

It is seen from (52) that in the absence of electric field ($\boldsymbol{E} = 0$) $dt_A = dt_B$, and at $p_{0\perp} = \tilde{p}_{0z} = 0$ the formula (39) is obtained, in which $a_0 = \dfrac{eE}{m_0}$.

## 5.  Gravitational field of a point particle

Let's consider the motion of a point particle of mass $m$ in gravitational field created by a point particle of mass $M$, being located at the origin of the reference frame $K$. The motion of particle is described by the equation (4), in which

$$\boldsymbol{F} = -\frac{\alpha\,\boldsymbol{r}}{r^3},\qquad(53)$$

where $\alpha = GmM$, $G$ is the gravitational constant.

By using the relation (5) connecting coordinates of particle in the reference frames $K$ and $\tilde{K}'$ with each other, transform the equation of motion (4) to the following equation describing the motion of particle in the $\tilde{K}'$-frame:

$$m\ddot{\boldsymbol{r}}' = -\frac{\alpha\bigl(\boldsymbol{r}' + \boldsymbol{r}_0(t)\bigr)}{\bigl|\boldsymbol{r}' + \boldsymbol{r}_0(t)\bigr|^3} - m\ddot{\boldsymbol{r}}_0(t) \equiv \tilde{\boldsymbol{F}}(\boldsymbol{r}',\,t).\qquad(54)$$

The requirement that the particle, being at rest at the origin of the reference frame $\tilde{K}'$, be free is expressed by equality

$$\tilde{\boldsymbol{F}}(0,\,t) = 0,\qquad(55)$$

which results in the equation defining the function $\boldsymbol{r}_0(t)$:

$$m\ddot{\boldsymbol{r}}_0(t) = -\frac{\alpha\,\boldsymbol{r}_0(t)}{r_0^3(t)}.\qquad(56)$$

Next, we expand the expression for the force $\tilde{\boldsymbol{F}}(\boldsymbol{r}',\,t)$ in a power series in $\boldsymbol{r}'$ and in this expansion retain, under the assumption that $\bigl|\boldsymbol{r}'\bigr| \ll r_0(t)$, only linear terms. As a result we obtain the following equation of motion:

$$m\ddot{\boldsymbol{r}}' = -\frac{\alpha}{r_0^3(t)}\left(\boldsymbol{r}' - \frac{3\,\boldsymbol{r}_0(t)\bigl(\boldsymbol{r}'\boldsymbol{r}_0(t)\bigr)}{r_0^2(t)}\right) \equiv \boldsymbol{F}_1(\boldsymbol{r}',t).\qquad(57)$$

As is seen from (57), the state of the particle, being at the origin of the reference frame $\tilde{K}'$, is an imponderability state. At $\boldsymbol{r}' \neq 0$ on the particle acts the force $\boldsymbol{F}_1(\boldsymbol{r}',t)$ resulting from the inhomogeneity of gravitational field.

Suppose that

$$\boldsymbol{r}_0(t) = r_0(t)\,\boldsymbol{n},\qquad(58)$$

where $0 < r_0(t) < \infty$, $\boldsymbol{n} = const$, $|\boldsymbol{n}| = 1$. Vector $\boldsymbol{n}$ characterizes the radial direction in which there takes place a free fall of the reference frame $\tilde{K}'$ on the force center, whose role is played by point particle of mass $M$. In this case equation (56) can be written in the form:

$$m\ddot{r}_0(t) = -\frac{\alpha}{r_0^2(t)}.\qquad(59)$$

Consider the peculiarities of the motion of particle in a vicinity of the origin of the reference frame $\tilde{K}'$. For simplicity, we neglect in equation (57) the dependence of quantity $r_0(t)$ upon $t$ considering this dependence to be weak enough ($r_0(t) = r_0$). Introduce the set of mutually orthogonal unit vectors of the reference frame $\tilde{K}'$, $\boldsymbol{e}_1, \boldsymbol{e}_2, \boldsymbol{e}_3$, such that $\boldsymbol{e}_3 = \boldsymbol{n}$, $[\boldsymbol{e}_3\boldsymbol{e}_1] = \boldsymbol{e}_2$, $[\boldsymbol{e}_1\boldsymbol{e}_2] = \boldsymbol{e}_3$. The solution of equation (57) is looked for in the form

$$\boldsymbol{r}' = \boldsymbol{e}_1 A_1 + \boldsymbol{e}_2 A_2 + \boldsymbol{e}_3 A_3,\qquad(60)$$

where $A_n = A_n(t)$ $(n = 1, 2, 3)$ are the required functions satisfying the set of equations

$$m\ddot{A}_1 = -\frac{\alpha}{r_0^3}A_1,\quad m\ddot{A}_2 = -\frac{\alpha}{r_0^3}A_2,\quad m\ddot{A}_3 = \frac{2\alpha}{r_0^3}A_3.\qquad(61)$$

The general solution of equations (61) is of the form:

$$A_n = A_{n0}\ cos(\ \omega_1 t + \varphi_n\ ), \quad n = 1, 2, \quad \omega_1 = \left( \frac{\alpha}{mr_0^3} \right)^{1/2},$$

$$A_3 = A_3'\ e^{-\omega_2 t} + A_3''\ e^{\omega_2 t}, \quad \omega_2 = \left( \frac{2\alpha}{mr_0^3} \right)^{1/2},$$

(62)

with $A_{n0}$, $\varphi_n$ $(n = 1, 2)$, $A_3'$ and $A_3''$ being arbitrary constants. According to (62), if we exclude from consideration the solution corresponding to unstable motion, i.e. if we choose the initial conditions so that $A_3'' = 0$, the picture of motion of the particle in the $\widetilde{K}'$-frame in a vicinity of point $r' = 0$ at $t \geq 0$ can be described as follows: along the $e_3$-axis the particle asymptotically approaches point $r' = 0$, and along the $e_1$ and $e_2$-axes it oscillates about this point with frequency $\omega_1$.

Note that the expression for the force $F_1(r', t)$, acting on particle in the reference frame $\widetilde{K}'$ (see (57) and (58)), depends on the choice of radial direction described by unit vector $n$, along which the free fall of the reference frame $\widetilde{K}'$ on the force center takes place. The existence of this dependence results in that the reference frames $\widetilde{K}'$, differing from each other only by the direction of translational motion relative to the reference system $K$ (i.e. by the free fall direction on force center), are physically nonequivalent to each other. There are, thus, infinitely large number of quasiinertial reference frames physically nonequivalent to each other, in each of which the imponderability state of particle is possible.

The motion of a particle, freely falling on force center along the direction $n$ in the reference frame $\widetilde{K}'$, is described by equation (54), where function $r_0(t)$ obeys equation (59). Let's proceed to a reference frame $\widetilde{\widetilde{K}}'$, which is obtained from the reference frame $\widetilde{K}'$ by way of its translation along the vector $n$ and which moves along this vector uniformly and rectilinearly relative to the reference frame $\widetilde{K}'$. The coordinates $(x'', y'', z'') \equiv r''$ and $(x', y', z') \equiv r'$ of particle in the above reference frames are connected with each other by equality $r' = r'' + a + bt$, where $a = an$, $b = bn$, $a, b = const$. Going from variables $r'$ to variables $r''$ in equation (54) and using notation

$$a + bt + r_0(t) = r_0'(t),$$

we obtain the following equation of motion:

$$m\ddot{r}'' = -\frac{\alpha(r'' + r_0'(t)n)}{\left| r'' + r_0'(t)n \right|^3} - m\ddot{r}_0'(t)n.$$

(63)

If the function $r_0'(t)$ obeys equation

$$m\ddot{r}_0'(t) = -\frac{\alpha}{r_0'^2(t)},$$

(64)

then equations (63) and (64) will coincide in form with equations (54) and (59), respectively. This means that the reference frames freely falling towards a force center in some radial direction are physically equivalent to each other if only they can be made coincident with each other by way of their translation in this direction and if, besides, they move uniformly and rectilinearly relative to each other.

Passing on to the solution of equation (59), we shall write first of all the energy conservation law resulting from it ($U(t)$ is potential energy):

$$\frac{mu_0^2(t)}{2} + U(t) \equiv E = const, \quad U(t) = -\frac{\alpha}{r_0(t)}.$$

(65)

Eliminating $r_0(t)$ from equalities (59) and (65), we come to the following equation of motion:

$$m \frac{du_0(t)}{dt} = -\frac{1}{\alpha} \left( \frac{mu_0^2(t)}{2} - E \right)^2. \qquad (66)$$

If the initial condition

$$u_0(t) = u_0 = const \quad \text{at} \quad t = 0 \qquad (67)$$

is imposed on the solution of equation (66), the latter proves to be equivalent to the following integral equation:

$$\int_{u_0}^{u_0(t)} \left( x^2 - \frac{2E}{m} \right)^{-2} dx = -\frac{m}{4\alpha} t. \qquad (68)$$

Consider the case, when the particle's falling on force center begins at the instant of time $t = 0$ with vanishing initial velocity: $u_0 = 0$. In virtue of (65)

$$r_0(0) = -\frac{\alpha}{E} \equiv r_0, \quad E < 0. \qquad (69)$$

In the region $\frac{mu_0^2(t)}{2} << |E|$ we expand the integrand in (68) in a series in powers of $\frac{m}{2E} x^2$ and restrict ourselves to several terms of expansion. As a result we obtain the expression:

$$u_0(t) - \frac{m}{3|E|} u_0^3(t) + \ldots = -t \frac{E^2}{\alpha m}.$$

From here we have:

$$u_0(t) = g_0 t + g_1 t^3 + \ldots, \quad r_0(t) = r_0 + \frac{1}{2} g_0 t^2 + \ldots,$$

$$g_0 = -\frac{\alpha}{mr_0^2}, \quad g_1 = -\frac{\alpha^2}{3m^2 r_0^5}. \qquad (70)$$

The proper time $d\tau$ of the particle, being at rest at the origin of the reference frame $\tilde{K}'$, can be calculated by formula (21) (at $dr' = 0$). Using the initial conditions mentioned above and taking into account the energy conservation law $\varepsilon_{kin}(t) + U(t) = U(0)$, we get:

$$\sqrt{1 - \frac{u_0^2(t)}{c^2}} = 1 + \frac{\alpha}{mc^2} \left( \frac{1}{r_0(0)} - \frac{1}{r_0(t)} \right). \qquad (71)$$

By introducing the potential of gravitational field

$$\varphi(t) = \frac{\alpha}{m} \left( \frac{1}{r_0(0)} - \frac{1}{r_0(t)} \right)$$

we arrive at the formula coincident in form with (43):

$$d\tau = \left( 1 + \frac{\varphi(t)}{c^2} \right) dt. \qquad (72)$$

From (72) the following expression is obtained for the relative course of time between points $A$ and $B$ in an inertial reference frame:

$$\frac{dt_A}{dt_B} = 1 - \frac{1}{c^2} \big( \varphi(t_A) - \varphi(t_B) \big) = 1 + \frac{\alpha}{mc^2} \left( \frac{1}{r_0(t_A)} - \frac{1}{r_0(t_B)} \right). \qquad (73)$$

Since the quantity $r_0(t)$ decreases as the particle approaches the force center (in this case according to (70) the magnitude of velocity $|u_0(t)|$ increases), by virtue of (72) and (73) the proper time of particle decreases, but the relative course of time in the inertial reference frame $K$ increases. Thus, in inertial reference frame, as the particle approaches the force center, time flows faster at the point where the particle is located. Note that in view of (70) potential $\varphi(t)$ can be written at $\left| r_0(t) - r_0(0) \right| << \left| r_0(0) \right|$ in the form $\varphi(t) = -g_0^2 t^2 / 2$, and therefore formula (73) for the relative course of time can be recast to the expression (40), in which one should make the substitution $a_0 \to g_0$.

## 6. Conclusion

The elucidation of the physical nature of time is one of the most important problems of theoretical physics. The purpose of research on the problem of time is to study the physical properties of time, i.e. to ascertain the possible interrelation between time and material processes. In particular, it is of interest to find out

- whether the flow of time depends upon physical processes and whether the back influence exists of the change of the course of time on physical processes,
- what mechanisms of the change of the course of time are available,
- what factors are capable to speed up or to slow down the flow of time.

In papers [5-8] on the basis of Lorentz transformations relating to coordinates of points, lying on the trajectory of motion of particle in a force field, the phenomenon of local dynamical inhomogeneity of time is predicted. The main result consists in the proof that material processes occurring in a physical system under the action of a force field necessarily influence the course of time along the trajectory of motion of particle. The case in point is the change of the course of time along particle's trajectory in one inertial reference frame as compared with that in the other.

In this paper the next step is made: the relationship is obtained which relates the course of time on one path section of a particle when moving in a force field to that on the other path section in the same inertial reference frame. The main idea underlying the approach developed results from the analysis of Lorentz transformations and consists in that the course of time of a particle moving by inertia, i.e. not subject to the influence of a force, should be uniform.

As is known [17,18], the existence of dependence of the course of time upon the gravitational field potential is predicted with the general theory of relativity (GTR). According to GTR ([17], p.303), time flows differently at the different points of space in one and the same reference frame. Since "gravitational field is nothing more nor less than a change of the space-time metric" ([17], p.313), one can assert, apparently, that the change in the course of time is due, in the view of GTR, to the change of the 4-space metric. It should be emphasized that in the present paper gravitational field is considered as an ordinary force field, and the particle motion is supposed to occur in pseudoeuclidian space-time. The main formulas of the article, (22) and (25), describe the change in the course of time in an arbitrary force field at different spatial points in one and the same inertial reference frame. As is seen from the results received, the change in the course of time in a force field is in no way connected with the change of space-time metric. It is conditioned by the force field action on particle in inertial reference frame and is a direct consequence of the dynamical principle underlying relativistic mechanics.

It should be emphasized that the existence of dependence of the course of time on the state of motion of particle in a force field points to the feasibility of controlling the course of time using force fields.

Note an important peculiarity of the noninertial reference frame in which the imponderability state of a particle is attained: there is such a space-time region in which the reference frame at hand can be approximately considered as inertial. In connection with the fact that such reference frames (it is natural to call them quasiinertial in contradistinction to the true inertial reference frames) are, generally speaking, not equivalent to each other (see previous section), the derivation of a rigorous criterion for inertial reference frame acquires especial significance. The dynamical criteria for defining the inertial and noninertial states are considered in the papers by B.I.Peschevitsky [19]. The heliocentric reference frame seems to be among the quasiinertial reference frames, being inertial with adequate accuracy only in a restricted region of space (for example, within the limits of our Galaxy) [16].

The author is grateful to Yu.D. Arepjev for his interest in the paper and stimulating discussions.